\documentclass[11pt]{article}
\usepackage{amsmath,amssymb,color,graphics,epsfig}

\usepackage{amsfonts}

\newcommand{\be}{\begin{equation}}
\newcommand{\ee}{\end{equation}}
\newcommand{\bea}{\setlength\arraycolsep{2pt} \begin{eqnarray}}
\newcommand{\eea}{\end{eqnarray}}
\newcommand{\nn}{\nonumber}

\def\ft#1#2{{\textstyle{\frac{\scriptstyle #1}{\scriptstyle #2} } }}
\def\fft#1#2{{\frac{#1}{#2}}}

\def\0{{\sst{(0)}}}
\def\1{{\sst{(1)}}}
\def\2{{\sst{(2)}}}
\def\3{{\sst{(3)}}}
\def\4{{\sst{(4)}}}
\def\5{{\sst{(5)}}}
\def\6{{\sst{(6)}}}
\def\7{{\sst{(7)}}}
\def\8{{\sst{(8)}}}
\def\sst#1{{\scriptscriptstyle #1}}

\begin{document}

\begin{flushright}
\end{flushright}

\vspace{25pt}
\begin{center}
{\large {\bf AdS and Lifshitz Scalar Hairy Black Holes in Gauss-Bonnet Gravity}}

\vspace{10pt}
 Bin Chen$^{1,2,3}$,\,Zhong-Ying Fan$^{1}$\,and Lu-Yao Zhu$^{2}$

\vspace{10pt}
{\it $^{1}$Center for High Energy Physics, Peking University, No.5 Yiheyuan Rd,\\}
{\it  Beijing 100871, P. R. China\\}
\smallskip
{\it $^{2}$Department of Physics and State Key Laboratory of Nuclear Physics and Technology,\\}
{\it Peking University, No.5 Yiheyuan Rd, Beijing 100871, P.R. China\\}
\smallskip
{\it $^{3}$Collaborative Innovation Center of Quantum Matter, No.5 Yiheyuan Rd,\\}
{\it  Beijing 100871, P. R. China\\}

\vspace{40pt}

\underline{ABSTRACT}
\end{center}
We consider Gauss-Bonnet (GB) gravity in general dimensions, which is non-minimally coupled to a scalar field. By choosing the scalar potential of the type $V(\phi)=2\Lambda_0+\fft 12m^2\phi^2+\gamma_4\phi^4$, we first obtain large classes of scalar hairy black holes
with spherical/hyperbolic/planar topologies that are asymptotic to locally anti-de Sitter (AdS) space-times. We derive the first law of black hole thermodynamics
using Wald formalism. In particular, for one class of the solutions, the scalar hair forms a thermodynamic conjugate with the graviton and nontrivially contributes to the
thermodynamical first law.  We observe that except for one
class of the planar black holes, all these solutions are constructed at the critical point of GB gravity where there exists an unique AdS vacua. In fact, Lifshitz vacuum is also allowed at the critical point. We then construct many new classes of neutral and charged Lifshitz
black hole solutions for a either minimally or non-minimally coupled scalar and derive the thermodynamical first laws. We also obtain new classes of exact dynamical
 AdS and Lifshitz solutions which describe radiating white holes. The solutions eventually become an AdS or
 Lifshitz vacua at late retarded times. However, for one class of the solutions the final state is an AdS space-time with a globally naked singularity.

\vfill {\footnotesize Emails: bchen01@pku.edu.cn, \quad fanzhy@pku.edu.cn, \quad lyzhu@pku.edu.cn}

\thispagestyle{empty}

\pagebreak

\tableofcontents
\addtocontents{toc}{\protect\setcounter{tocdepth}{2}}




\section{Introduction}
Recently, there are large classes of scalar hairy black holes that are asymptotic to Minkowski
\cite{Anabalon:2013qua,Anabalon:2012dw,Gonzalez:2013aca,Feng:2013tza,Fan:2015oca} and anti-de Sitter (A)dS space-times
\cite{Anabalon:2012dw,Gonzalez:2013aca,Feng:2013tza,Fan:2015oca,Henneaux:2002wm,Martinez:2004nb,Anabalon:2013sra,Acena:2013jya,Zloshchastiev:2004ny,Fan:2015tua,
Fan:2015ykb,Ayon-Beato:2015ada}
 having been analytically constructed in Einstein gravity either minimally or non-minimally coupled to a scalar field. The solutions provide a lot of counter
 examples for the no-hair theorems which exclude the existence of hairy black holes in General Relativity. It was shown in \cite{Lu:2014maa} that for the general
 two parameter family solutions in Einstein-Scalar gravity, the first law of black hole thermodynamics can be modified by a one form associated with the scalar. This is straightforwardly confirmed by numerical solutions.
 However, for all the analytical solutions constructed above, the thermodynamical first law does
 not receive any contribution from the scalar because either the scalar contains only one normalizable mode at asymptotic infinity or the two independent fall-off modes ($\psi_1\,,\psi_2$) of the scalar at infinity satisfy a relation $\psi_2^{\Delta_1}\propto \psi_1^{\Delta_2}$, where $\Delta_1\,,\Delta_2$ are the scaling dimensions of $\psi_1\,,\psi_2$ respectively (The technique details are given in our Appendix). The only known examples that have a nontrivial scalar contribution to the thermodynamical first law are the Kaluza-Klein AdS dyonic
 black hole and its multi-charge generalizations in maximal gauged supergravity \cite{Lu:2013ura,Chow:2013gba}. Thus it is interesting to construct more analytical solutions
whose first law of thermodynamics are modified by the scalar. On the other hand, for some classes of the static hairy black hole solutions,
one can promote the ``scalar charge" to be dependent on the advanced/retarded time and obtain exact dynamical solutions in Eddington-Finkelstein-like coordinates \cite{Fan:2015tua,Fan:2015ykb,Ayon-Beato:2015ada,Fan:2016yqv,Zhang:2014sta,Zhang:2014dfa,Lu:2014eta,Xu:2014xqa}.
These solutions provide analytical examples for black holes formation and may have further applications in the AdS/CFT correspondence.

The motivation of current paper is to construct more scalar hairy black hole solutions in Gauss-Bonnet (GB) gravity which is non-minimally coupled to a scalar.
 It is well known that GB gravity is of second order derivative and ghost free. It is a proper generalization of Einstein gravity with quadratic corrections
in dimensions higher than four. However, due to the existence of higher curvature terms, the equations of motion become highly nonlinear even if for the specialized static
 ansatz ($-g_{tt}=g^{rr}$) that we consider.
 This is the main difficulty that we encounter in constructing new black hole solutions with scalar hair. Fortunately, it turns out that the equations of motion become easily
 solvable for the critical GB coupling where there exists an unique (A)dS vacua\footnote{Depending on the parameters and assuming the bare cosmological constant is negative
 $\Lambda_0<0$, there are either two distinct AdS vacua or one AdS and one dS. For positive $\Lambda_0$, there will be either two distinct dS vacua or one dS and one
 AdS. For vanishing $\Lambda_0$, there will be one Minkowski vacua and one (A)dS vacua. }. In fact, Lifshitz vacua is also allowed at the critical point
 \cite{Dehghani:2010kd,Dehghani:2010gn}. We then construct large classes of AdS and Lifshitz black hole solutions with scalar hair at the critical point. The scalar potential can be unified expressed as $V=2\Lambda_0+\fft 12 m^2\phi^2+\gamma_4 \phi^4$, which is much
 simpler than those constructed in Einstein gravity. For generic GB coupling, we also obtain one class of AdS planar black holes. We study the global properties of
 the solutions and adopt Wald formalism to derive the thermodynamical first laws. In particular, one class of the solutions
 contains two independent integration constants which are associated with the scalar
and the graviton mode respectively. Although the scalar has only one normalizable mode at infinity, ``the scalar charge" forms a thermodynamic conjugate with the graviton mode
and non-trivially contributes to the thermodynamical first laws. Finally, for some classes of the solutions, we promote
 the ``scalar charge" to be dependent on the retarded time and obtain exact dynamical AdS and Lifshitz solutions describing radiating white holes .
The solutions eventually become an AdS/Lifshitz vacua at late retarded times but one class of the solutions approaches a static AdS space-time with a naked singularity.

The paper is organized as follows. In section 2, we consider GB gravity non-minimally coupled to a scalar and derive the covariant equations of motion.
In section 3, we briefly review the
Wald formalism and derive various quantities of the Wald formula for our gravity model. We also provide more explicit formulas for general static solutions.
In section 4, we demonstrate our construction method and construct large classes of spherical/hyperbolic/planar black holes with scalar hair that are asymptotical to locally AdS space-times. We study the global properties and derive the first law of thermodynamics using Wald formalism.
In section 5, we obtain many new classes of neutral and charged Lifshitz black holes and derive the thermodynamical first laws. Finally, we
conclude this paper in section 6.

\section{GB gravity with a non-minimally coupled scalar}
We consider Gauss-Bonnet (GB) gravity non-minimally coupled to a scalar field, together with a generic scalar potential. The Lagrangian density is given by
\be \mathcal{L}=R+\alpha \big(R^2-4R_{\mu\nu}^2+R_{\mu\nu\rho\sigma}^2\big)-\fft 12 (\partial \phi)^2-\fft 12 \eta R \phi^2-V(\phi)\,. \label{lagrangian}\ee
where $\alpha$ is the Gauss-Bonnet coupling constant and $\eta$ is a constant characterizing the coupling strength between the scalar and the curvature. Note that the GB term is also non-minimally coupled to the scalar field in string generated models \cite{Boulware:1985wk,Zangeneh:2015tva} but for our purpose we shall not consider this more complicated case.
The effective Newton's ``constant" now becomes inversely proportional to
\be \kappa(\phi)\equiv 1-\ft 12 \eta \phi^2 \,.\label{kappa}\ee
which is dependent on the scalar field. In order to avoid ghost-like graviton modes, we shall require $\kappa(\phi)$ being positive throughout this paper.
The covariant equations of motion are
\be
 G_{\mu\nu}+\alpha H_{\mu\nu}=T_{\mu\nu}^{\rm (min)}+T_{\mu\nu}^{\rm (non)}\,,\qquad
\Box\phi = \eta\phi R + \fft{dV}{d\phi}\,,\label{non-min-geneom}
\ee
where $ G_{\mu\nu}=R_{\mu\nu}-\fft 12 R g_{\mu\nu}$ is the Einstein tensor and
\be H_{\mu\nu}=2\Big(R R_{\mu\nu}-2R_{\mu\rho}R^\rho_{\nu}-2R^{\tau\sigma}R_{\tau\mu\sigma\nu}+R_{\mu\rho\tau\sigma}R_{\nu}^{\ \rho\tau\sigma} \Big)
               -\fft 12 g_{\mu\nu}\Big(R^2-4R_{\tau\sigma}^2+R_{\lambda\rho\tau\sigma}^2 \Big)\,.
               \ee
The energy-momentum tensors are given by
\bea
T_{\mu\nu}^{\rm (min)}&=&\fft 12\partial_\mu\phi \partial_\nu \phi-\fft 12 g_{\mu\nu}\Big(\ft 12(\partial \phi)^2+V(\phi) \Big)\,,\nn\\
T_{\mu\nu}^{\rm (non)}&=&\fft12\eta(\phi^2 G_{\mu\nu}+g_{\mu\nu}\Box \phi^2-\nabla_\mu \nabla_\nu \phi^2)\,.
\eea
 This model was previously studied in \cite{Gaete:2013ixa,Gaete:2013oda} for a certain type potential.
 For vanishing GB coupling $\alpha$, the model is well studied in \cite{Fan:2015tua}, where large classes of static and dynamic hairy planar black holes have been analytically constructed. In current paper, we will try to construct new scalar hairy black hole solutions in the model with non-vanishing GB coupling. We only consider static solutions with spherical/hyperbolic/toric isometries. The most general ansatz takes the form of
\be ds^2=-hdt^2+dr^2/f+r^2d\Omega_{n-2,k}^2 \,,\qquad \phi=\phi(r)\,,\label{ansatz}\ee
where $h\,,f$ are functions of $r$, $d\Omega_{n-2,k}^2$ denotes the metric of the $(n-2)$ dimensional space with constant curvature $k=0\,,\pm 1$.
 In order to construct exact black hole solutions as many as possible, we will also introduce an additional Maxwell field which is minimally coupled to the gravity and the scalar when necessary.



\section{Wald formalism and black hole thermodynamics}

\subsection{A brief review}
Wald formalism provides a systematic procedure for deriving the first law of black hole thermodynamics for a generic (local and covariant) gravity theory.
It was first demonstrated in \cite{wald1,wald2}. Let us first gives a brief review in the following.

Variation of the action with respect to the metric and matter fields, one finds
\be S=\int \mathrm{d}^nx\sqrt{-g}\,\mathcal{L}[\psi] \,,\qquad \delta S=\int \mathrm{d}^nx\sqrt{-g}\big(E_\psi \delta \psi+\triangledown_\mu J^\mu \big)\,,\ee
where $\psi$ collectively denotes the dynamical fields and $E_\psi=0$ are the equations of motion.
Given the current $J^\mu$,  one can define a current 1-form and its Hodge dual
\be
J_\1=J_\mu dx^\mu\,,\qquad
\Theta_{\sst{(n-1)}}={*J_\1}\,.
\ee
When the variation is generated by an arbitrary vector field $\xi$, one can define an associated Noether current $(n-1)$-form as
\be
J_{\sst{(n-1)}}=\Theta_{\sst{(n-1)}}-i_\xi\cdot {*\mathcal{L}}\,,
\ee
where $i_\xi\cdot$ denotes the contraction of $\xi$ with the first index of the differential form it acted upon. It was shown in \cite{wald1,wald2} that the Noether
current $J_{(n-1)}$ is closed once the equations of motion are satisfied
\be
dJ_{\sst{(n-1)}}= 0.
\ee
Thus a Noether charge $(n-2)$-form can be defined by
\be J_{\sst{(n-1)}}= dQ_{\sst{(n-2)}}\,.
\ee
 It was shown in\cite{wald1,wald2} that when $\xi$ is a Killing vector field, the variation of the Hamiltonian with respect to the integration constants of a given solution is:
\be
\delta H=\frac{1}{16\pi}\Big[\delta \int_{\mathcal{C}}J_{\sst{(n-1})}- \int_{\mathcal{C}}d(i_\xi\cdot \Theta_{\sst{(n-2)}})\Big]=\frac{1}{16\pi}\int_{\Sigma_{n-2}}\Big[\delta Q_{\sst{(n-2)}}-i_\xi\cdot \Theta_{\sst{(n-2)}}\Big]\,.\label{generalwald}
\ee
where $\mathcal{C}$ is a Cauchy surface and $\Sigma_{n-2}$ is its two boundaries, one on the horizon and the other at infinity. For a stationary black hole, the
Killing vector $\xi$ becomes null on the horizon and the thermodynamical first law emerges by the vanishing of $\delta H$. Hence evaluating $\delta H$ at the boundaries yields
\be
\delta H_\infty = \delta H_+\,.
\ee
On the horizon, one has
\be \delta H_+=T \delta S \,,\label{waldhorizon}\ee
where the temperature and the entropy are given by
\be
T=\fft{\kappa}{2\pi}\,,\qquad S=-\fft{1}{8} \int_+ \sqrt{\gamma}\, d^{n-2}x\, \epsilon_{ab}\epsilon_{cd}\, \fft{\partial \mathcal{L}}{\partial R_{abcd}}\,.
\ee
Here $\kappa$ is the surface gravity on the horizon. It should not be confused with the effective gravitational coupling ``constant" in (\ref{kappa}). The thermodynamical first law of a black hole is simply
\be \delta H_{\infty}=T \delta S   \,.\ee
However, there is no general formula for the evaluation of $\delta H_{\infty}$ at asymptotic infinity and one needs study it in the case-by-case basis.
Some interesting examples can be found in the literature \cite{Lu:2014maa,Liu:2013gja,Liu:2014tra,Fan:2014ixa,Fan:2014ala,Liu:2014dva,Liu:2015tqa}.

\subsection{Applying for GB gravity}
Various quantities in the Wald formula (\ref{generalwald}) have been explicitly given in \cite{Fan:2014ala} for Einstein gravity extended with
quadratical curvature invariants. GB gravity is also included as a special case. We have (for pure GB gravity)
\bea
J_{\sst{(n-1)}}^{(G)} &=&2\varepsilon_{\mu c_1...c_{n-1}} \nabla_\lambda\Big(\nabla^{[\lambda}\xi^{\mu]}
+2\alpha\big(R \nabla^{[\lambda}\xi^{\mu]}-4R^{\sigma[\lambda}\nabla_\sigma \xi^{\mu]}+ R^{\lambda\mu\sigma\rho}\nabla_\sigma \xi_\rho\big)\Big)  \,,\cr
Q_{\sst{(n-2)}}^{(G)} &=& -\varepsilon_{\mu\nu c_1...c_{n-2}}\Big(\nabla^{\mu}\xi^{\nu}+
+2\alpha\big(R\nabla^{\mu}\xi^{\nu}-4 R^{\sigma\mu}\nabla_\sigma \xi^{\nu}+ R^{\mu\nu\sigma\rho}\nabla_\sigma \xi_\rho \big)\Big)\,,\cr
i_{\xi}\cdot \Theta_{(n-1)}^{(G)}&=&-\varepsilon_{\sigma\mu c_1...c_{n-2}}\xi^{\sigma}\Big(
\big(G^{\mu\nu\rho\lambda}+2\alpha (R G^{\mu\nu\rho\lambda}-2 T^{\mu\nu\rho\lambda}+2 R^{\mu\rho\lambda\nu} )\big) \nabla_\nu \delta g_{\rho\lambda}\cr
&&-2\alpha \big(G^{\mu\nu\rho\lambda}\nabla_\nu R-2 \nabla_\nu T^{\nu\mu\rho\lambda}+2 \nabla_\nu R^{\mu\rho\lambda\nu}\big) \delta g_{\rho\lambda}\Big)\,,\label{various}
\eea
where some quantities are defined by
\bea G^{\mu\nu\rho\sigma} &\equiv&\ft 12 (g^{\mu\rho}g^{\nu\sigma}+g^{\mu\sigma}g^{\nu\rho}) - g^{\mu\nu}g^{\rho\sigma},\cr
T^{\mu\nu\rho\lambda} &\equiv& g^{\mu\rho}R^{\nu\lambda} +g^{\mu\lambda}R^{\nu\rho}-g^{\mu\nu}R^{\rho\lambda}-
g^{\rho\lambda}R^{\mu\nu}\,.\label{gt}
\eea
Note that we have put the non-minimal coupling term into the scalar sector. By simple calculations, we find
\bea J_{\sst{(n-1)}}^{(\phi)} &=&\eta\,\varepsilon_{\mu c_1...c_{n-1}} \nabla_\lambda
\Big(2\nabla^{[\lambda}\phi^2\xi^{\mu]}-\phi^2\nabla^{[\lambda}\xi^{\mu]} \Big)  \,,\cr
Q_{\sst{(n-2)}}^{(\phi)} &=& -\fft 12 \eta\, \varepsilon_{\mu\nu c_1...c_{n-2}}\Big( 2\nabla^{\mu}\phi^2\xi^{\nu}-\phi^2\nabla^{\mu}\xi^{\nu}  \Big)\,,\cr
i_{\xi}\cdot \Theta_{(n-1)}^{(\phi)}&=&\varepsilon_{\sigma\mu c_1...c_{n-2}}\xi^{\sigma}\Big(\nabla^\mu \phi \delta \phi-
\ft 12 \eta\, G^{\mu\nu\rho\lambda}\big(\nabla_\nu\phi^2-\phi^2\nabla_\nu \big)\delta g_{\rho\lambda}   \Big)\,.\label{various2}
\eea
For the general static solutions (\ref{ansatz}), let $\xi=\partial/\partial t$, we obtain
\be
\delta H^{(G)}= \fft{\omega}{16\pi} r^{n-2}\, \sqrt{\fft{h}{f}}\, \Big(
-\fft{n-2}{r} + \fft{2\alpha(n-2)(n-3)(n-4) (f-k)}{r^3} \Big)\delta f\,,
\label{hgrav}\ee
and
\bea
  &&\delta  H^{(\phi)}=\fft{\omega}{16\pi} r^{n-2}\, \sqrt{h f}\big(-\phi'\delta \phi+\eta \phi \Delta_\phi  \big)\,,\nn\\
&&\Delta_\phi=2\delta\phi'+\Big(\fft{2\phi'}{\phi}-\fft{h'}{h}
 \Big )\delta \phi+\fft{\phi}{f}\Big(\fft{\phi'}{\phi}+\fft{n-2}{2r} \Big)\delta f \,.\label{hphi}
\eea
Here $\omega$ is the volume factor of the $(n-2)$ dimensional space. The total variation of the Hamiltonian is given by the sum of above two terms.

Now let us verify the identity (\ref{waldhorizon}) on the horizon
for our gravity model. We develop Taylor expansions for the metric functions and the scalar field in the near horizon region,
\bea
&&h=h_1(r-r_0)+h_2(r-r_0)^2+h_3(r-r_0)^3+\cdots\,,\nn\\
&&f=f_1(r-r_0)+f_2(r-r_0)^2+f_3(r-r_0)^3+\cdots\,,\nn\\
&&\phi=\phi_0+\phi_1(r-r_0)+\phi_2(r-r_0)^2+\phi_3(r-r_0)^3+\cdots\,.
\label{horizon}\eea
By plugging the expansions into the equations of motion, we find that there are three independent parameters on the horizon which we may take to be ($r_0\,,\phi_0\,,h_1$).
The rest coefficients are determined in terms of these three parameters. Note that $h_1$
 is a trivial parameter associated with the scaling symmetry of the time coordinate and can be set to unity.
Substituting these expansions into the Wald formula, we obatin
\be \delta H_+=\fft{\omega}{16\pi}r_0^{n-3}\sqrt{h_1f_1}\Big((n-2)\Big(1-\ft 12 \eta \phi_0^2+\fft{2\alpha k (n-3)(n-4)}{r_0^2} \Big)\delta r_0-\eta r_0
\phi_0\delta\phi_0 \Big)\,,\ee
The temperature and entropy are given  by
\be T=\fft{1}{4\pi}\sqrt{h_1f_1}\,,\qquad S=\fft{1}{4}\omega r_0^{n-2}\Big(1-\ft 12\eta \phi_0^2+\fft{2\alpha k(n-2)(n-3)}{r_0^2} \Big)\,. \ee
It is straightforward to verify that (\ref{waldhorizon}) is indeed satisfied.

In asymptotically AdS  space-times, the scalar has two independent fall-offs at infinity
\be \phi=\fft{\psi_1}{r^{\fft{n-1-\sigma}{2}}}+\fft{\psi_2}{r^{\fft{n-1+\sigma}{2}}}+\cdots \,,\label{fallscalar1}\ee
where $\sigma$ is defined by
\be \sigma=\sqrt{(n-1)\Big((1-4\eta)n-1\Big)+4m^2\ell^2} \,.\ee
 Note that the Breitenlohner-Freedman (BF) bound has been shifted by the nonminimal coupling constant, given by $m^2_{BF}=-\fft{1}{4} (n-1)\Big((1-4\eta)n-1\Big)g^2$,
 where $g$ denotes the inverse of the effective AdS radius $g=1/\ell$. The asymptotic behaviors of the metric functions strongly depend on the GB coupling $\alpha$.
We find that
 \be h=g^2 r^2+k-\fft{\mu_g}{r^{n-3}}+\cdots\,,\qquad f=g^2 r^2+k+\cdots \,,\label{metricfall1}\ee
  for generic GB coupling $\alpha$ and\footnote{The linearized equations of motion are automatically cancelled at the critical point so that the fall-off of the graviton mode is determined by the quadratic equations of motion. This is why the metric functions behave different at the critical point. This fact plays an important role in studying the global properties of black hole solutions.}
 \be h=g^2 r^2+k-\fft{\mu_g}{r^{\fft{n-5}{2}}}+\cdots\,,\qquad f=g^2 r^2+k+\cdots \,,\label{metricfall2}\ee
for the critical GB coupling $\alpha=\alpha_c=1/\big(2(n-3)(n-4)g^2\big)$, where the AdS vacua becomes unique. It should be emphasized that there could be slower fall-offs than those displayed in the large-r expansions (\ref{fallscalar1}\,,\ref{metricfall1}\,,\ref{metricfall2}) for general scalar mass due to back-reaction effects. By substituting into the equations of motion, it is
easy to see that there are three independent parameters $(\mu_g\,,\psi_1\,,\psi_2)$ at asymptotic infinity. All the higher order coefficients can be solved in terms
of these three parameters. However, integrating out to infinity from the near horizon solutions (\ref{horizon}), we find that the three parameters at asymptotic infinity are determined by functions of the two non-trivial
parameters $(r_0\,,\phi_0)$ on the horizon. Hence, the general static black hole solutions are characterized by only two independent parameters.
The existence of two parameter family black hole solutions is straightforwardly confirmed by numerical calculations. We also obtain one class of analytical black hole solutions with
two independent parameters (see sec \ref{exception} for detail).
\subsection{Definition of black hole mass}
In the remaining of this paper, we will apply the explicit Wald formulas Eq.(\ref{hgrav}-\ref{hphi}) to derive the thermodynamical first laws of scalar hairy black holes. One of key quantities is the
energy or black hole mass. For pure GB graity, $\delta H_\infty$ is integrable and one can define the black hole mass as $\delta M\equiv \delta H_\infty$. However, in the
presence of a scalar, $\delta H_\infty$ is in general non-integrable. It was shown in \cite{Lu:2014maa} that for the general two parameter family black hole solutions in Einstein-Scalar gravity, one finds
\be \delta H_\infty=\delta M_{\mathrm{ther}}+K(\delta\psi_1\,,\delta\psi_2) \,,\label{genelaw}\ee
where $\delta M_{\mathrm{ther}}$ is an integrable part associated with the graviton mode, $K(\delta\psi_1\,,\delta\psi_2)$ is the non-integrable part which is a
one form, given by the linear combination of $\delta \psi_1$ and $\delta \psi_2$. We find that generally the relation (\ref{genelaw}) is also valid for our gravity model.
 The specific form of $K(\delta\psi_1\,,\delta\psi_2)$ strongly
depends on the full large-r expansion of the solutions which is rather involved. To be concrete, we give an explicit example in the Appendix. The detail is
irrelevant in our discussion. The non-integrability of $\delta H_\infty$ may be interpreted as that
the solution have no well-defined mass \cite{Chow:2013gba} but we prefer a different interpretation \cite{Lu:2014maa,Liu:2015tqa} that the relation (\ref{genelaw})
 provides a definition of the ``thermodynamic mass" $M_{\mathrm{ther}}$, which has the dimension of the energy and becomes the ``true" black hole mass in the absence
 of the scalar.

 For the exact black hole solutions that we constructed in this paper, the scalar field contains only one normalizable mode at asymptotic infinity such that the scalar
 one form $K(\delta \psi_1\,,\delta \psi_2)$ in general vanishes. Thus for our solutions, $\delta H_{\infty}$ becomes integrable and
 $\delta M_{\mathrm{ther}}=\delta H_\infty$, implying that the ``thermodynamic mass" coincides with the black hole mass.
 Without confusion, we will simply refer the ``thermodynamic mass" as the black hole mass in the following sections.

Finally, we remark that above discussions and the relation (\ref{genelaw}) are invalid for the critical coupling constants $\alpha=\alpha_c$ and $\eta=\fft{n-1}{4(n+1)}$.
Actually we obtain the exact black hole solutions with two independent parameters. We find that although the scalar also has only one normalizable mode at asymptotic infinity, ``the scalar charge" forms a
thermodynamic conjugate with the graviton mode and contributes to the corresponding first law. For detail, see sec \ref{exception}.

\section{AdS black holes}

In this section, we will construct AdS black hole solutions with scalar hair in our gravity model. We consider a specialized ansatz
\be ds^2=-fdt^2+dr^2/f+r^2 d\Omega_{n-2,k}^2 \,,\qquad \phi=\phi(r)\,.\ee
It turns out that the scalar satisfies a decoupled equation
\be (2\eta-1)\phi'^2+2\eta \phi\phi''=0 \,,\ee
which is obtained by $E^t_{\,\,\,t}-E^r_{\,\,\,r}$. The equation can be immediately solved by
\be \phi=\fft{1}{(c_1 r+c_2)^\mu}\,,\qquad \eta=\fft{\mu}{2(2\mu+1)}  \,,\label{phi}\ee
where $c_1$, $c_2$ are two integration constants and $\eta$ has been reparameterized by $\mu$. The scalar potential can be solved as a function of $r$ from
the equations of motion. Then the remaining equation of motion is a second order nonlinear differential equation for the metric function $f$ (the nonlinear terms come from the GB higher curvature interaction). If this equation can also be solved exactly, the scalar potential can be further expressed as a function of $\phi$ according to the relation (\ref{phi}). This is a general procedure that we apply to construct new black holes solutions with scalar hair. The main difficulty that we encounter is how to exactly solve the metric function $f$. For Einstein gravity the situation becomes simple and large classes of static and dynamic hairy planar black holes have been reported in \cite{Fan:2015tua}. When the GB term was turned on, we shall set $c_2=0$ for the construction to be as simple as possible. We find that the non-linear differential equation of the metric function $f$ can be easily solved at the critical GB coupling $\alpha=\alpha_c$. We then successfully construct many new classes of AdS scalar hairy black hole solutions at the critical point. We also obtain one class of planar black hole solutions for generic GB coupling.

In the remaining of this paper, we will not repeat the construction details but simply provide the final results.

\subsection{Spherical black hole in $n=5$ dimension}

\subsubsection{Static solution}
Let us first study spherical black holes with scalar hair. We consider a massless scalar with the potential given by
\be V=-6g^2-\fft{27}{1024}g^2\phi^4 \,.\ee
where $g$ denotes the inverse of the effective AdS radius $g=1/\ell$. We find that when the Gauss-Bonnet coupling
 takes the critical value $\alpha=\alpha_c=1/(4g^2)$, there exists an exact spherical black hole solution with scalar hair for $\eta=3/16$
\bea &&ds^2=-fdt^2+dr^2/f+r^2 d\Omega_{3}^2 \,, \nn\\
     &&\phi=\Big( \fft{q}{r}\Big)^{3/2}\,,\quad f=g^2 r^2+3-\fft{3g^2q^3}{32 r} \,. \label{sol1}\eea
Here $q$ is the only independent integration constant associated with the scalar. Note that the BF bound is satisfied because $m_{BF}^2=-\ft 14 g^2$ for $\eta=3/16$.
For vanishing $q$, the solution becomes
\be ds^2=-\big( g^2r^2+3 \big)dt^2+\fft{dr^2}{\big(g^2r^2+3\big)}+r^2d\Omega^2_3  \,,\qquad \phi=0\,,\label{naked}\ee
which however is not an AdS vacua. In fact, the solution (\ref{naked}) belongs to a special type solutions of GB gravity with a single AdS vacua \cite{Crisostomo:2000bb,Aros:2000ij,Fan:2016zfs}. Note that the space-time contains a naked singularity at the origin. For non-vanishing scalar, the singularity is dressed by an event horizon which is determined by the foot of $f(r_0)=0$.
Evaluating $\delta H$ at infinity, we find
\be \delta H_\infty=0 \,,\ee
which implies that the black hole mass vanishes. We find that the
entropy also vanishes
\be S=0\,,\ee
though the temperature is non-vanishing, given by
\be T=\fft{2(g^2r_0^2+1)}{4\pi r_0} \,. \ee
 Thus the thermodynamical first law is trivially satisfied. Note that the black hole arrives at its ground state at a finite but nonzero temperature. This is characteristic for our solution. It deserves a deeper understanding since one in general does not expect to observe this in a non-gravitational thermodynamic system.

 Finally, we shall point out that the solution is only valid in $n=5$ dimension and cannot be generalized to higher ($n\geq 6$) dimensions. This is left as an open problem.

\subsubsection{Dynamical solution}
Now we try to search new dynamical solutions. We first rewrite the static solution (\ref{sol1}) in Eddington-Finkelstein-like coordinates.
The key point is the integration constant $q$ associated with the scalar is a non-conserved quantity. We can promote it to be dependent
on the advanced/retarded time coordinate and solve all the equations of motion. We obtain
\bea &&ds^2=-fdu^2-2dudr+r^2d\Omega_3^2\,,\nn\\
     &&\phi=\Big( \fft{a}{r}\Big)^{3/2}\,,\quad f=g^2 r^2+3-\fft{3g^2a^3}{32 r}\,,
\eea
where $a\equiv a(u)$ is a single function of the retarded time $u$ and satisfies a second order nonlinear differential equation
\be a \ddot a-2\dot a^2=0 \,.\label{aeq}\ee
Here a dot denotes the derivative with respect to the retarded time $u$. The equation can be solved by
\be a=\fft{1}{\tilde{c}_1 u+\tilde{c}_2} \,,\ee
where $\tilde{c}_1\,,\tilde{c}_2$ are two integration constants. For $\tilde{c}_1=0$, we rediscover the static black hole solution (\ref{sol1}). For $\tilde{c}_1\neq 0$, we obtain a new dynamical
solution. For convenience, we set $\tilde{c}_2=0$ and require $\tilde{c}_1>0$ to let the solution characterizing a dynamical white hole at $u>0$ region. Then the solution describes a radiating white hole that eventually becomes the static AdS space-time (\ref{naked}) with a naked singularity at late retarded times.

\subsection{Planar black holes in general dimensions}
 Now we construct new classes of static hairy planar black holes that are asymptotical to AdS space-times in general dimensions. The solutions can be  classified into different classes according to the form of the potential.
\subsubsection{Case 1: with massless scalar}
The first class solution we present is valid for a massless scalar. The potential is given by
\be V=-(n-2)(n-1)g^2\Big(1-(n-4)(n-3)\alpha g^2\Big)-\fft{(n-1)^3\phi^4}{256\alpha n(n-4)(n-3)}  \,,\ee
For generic GB coupling, we obtain
\bea \label{ps1}
&&ds^2=-fdt^2+dr^2/f+r^2 dx^idx^i \,,\nn\\
&&\phi=\Big(\fft{q}{r}\Big)^{\fft{n-1}{2}}\,,\quad f=g^2 r^2-\fft{(n-1)q^{n-1}}{16\alpha n (n-4)(n-3)r^{n-3}}\,,
\eea
where the non-minimal coupling constant $\eta$ is given by (\ref{phi}) with $\mu=(n-1)/2$. The scalar satisfies the BF bound because now $m_{BF}^2=0$.
 The existence of the event horizon requires the GB coupling $\alpha$ being positive. Interestingly, the metric function $f$ takes a Schwarzschild-like form and
one may expect that the mass of the black holes is linearly proportional to $q^{n-1}$. Evaluating $\delta H_\infty$ yields
\be \delta H_\infty=\fft{(n-2)(n-1)^2\omega q^{n-2}\delta q }{256\pi\alpha n(n-3)(n-4)}\Big(1-2(n-3)(n-4)\alpha g^2 \Big) \,,\ee
which implies that  the black hole mass is given by
\be\label{mass1} M=\fft{(n-2)(n-1)\omega q^{n-1} }{256\pi\alpha n(n-3)(n-4)}\Big(1-2(n-3)(n-4)\alpha g^2 \Big)   \,.\ee
This is consistent with our naive expectation. In fact, the mass (\ref{mass1}) is simply the AMD mass \cite{Ashtekar:1984zz,Ashtekar:1999jx} generalized in higher curvature gravity \cite{Okuyama:2005fg,Pang:2011cs}.
The temperature and entropy are given by
\be T=\fft{(n-1)g^2 r_0}{4\pi}\,,\qquad S=\fft 14\omega r^{n-2}_0 \Big(1-2(n-3)(n-4)\alpha g^2 \Big)  \,.\ee
It follows that the thermodynamical first law
\be dM=T dS \,,\ee
and the Smarr-like relation
\be M=\fft{n-2}{n-1}T S \,,\ee
are satisfied. We shall point out that for AdS planar black holes, there exists an additional scaling symmetry, which leads to above Smarr-like relation \cite{Liu:2015tqa}.

\subsubsection{Case 2: with massive scalar}\label{exception}
The second class solution we present is valid for a massive scalar. The potential is given by
\be V=-\fft 12(n-2)(n-1)g^2+\fft{(n-3)(n-1)^2}{32(n+1)}g^2\phi^2  \,.\ee
The mass square of the scalar is positive definite, given by $m^2=\fft{(n-3)(n-1)^2}{16(n+1)}g^2 $. The
Breitenlohner-Freedman (BF) bound is always satisfied for the non-minimal coupling $\eta=(n-1)/\big(4(n+1)\big)$, where we find the black hole solutions because $m^2=m_{BF}^2+\ft{1}{16}(n-1)^2g^2$. For the critical GB coupling
$\alpha=\alpha_c$, we obtain the exact black hole solutions with two independent parameters
\bea\label{ps2}
&&ds^2=-fdt^2+dr^2/f+r^2 dx^idx^i \,,\nn\\
&& \phi=\Big(\fft{q}{r}\Big)^{\fft{n-1}{4}}\,,\qquad f=g^2r^2-\fft{\mu_g}{r^{\fft{n-5}{2}}} \,.
\eea
where $q$ and $\mu_g$ are the two independent integration constants associated with the scalar and the graviton mode respectively. This solution was first studied in \cite{Gaete:2013ixa} as ``stealth solutions". The equations of motion are satisfied as $E_{\mu\nu}=0=T_{\mu\nu}$. For vanishing scalar, the solutions are also valid for spherical/hyperbolic topologies. By plugging the solutions into the Wald formula Eq.(\ref{hgrav}-\ref{hphi}), we find
\be\label{nonh} \delta H_\infty=\fft{(n-2)\omega }{16\pi g^2}\mu_g \delta \mu_g-\fft{(n-1)\omega }{256\pi (n+1)}\Big((n-3)\psi \delta \mu_g+ (n-1) \mu_g \delta \psi \Big)\,, \ee
where $\psi\equiv q^{\fft{n-1}{2}}$. It is clear that $\delta H_{\infty}$ receives contributions from the scalar and becomes non-integrable. Surprisingly, the general relation (\ref{genelaw}) becomes invalid for this solution. The one form associated with the scalar is now determined by both the scalar and the graviton mode. Consequently, the scalar forms a thermodynamic conjugate with the graviton mode and contributes to first law of thermodynamics, although it contains only one normalizable mode at asymptotic infinity.
The ``thermodynamic mass" can be defined by
\be M_{\mathrm{ther}}=\fft{(n-2)\omega}{32\pi g^2}\mu_g^2\,.\ee
The corresponding first law can be expressed as
\be\label{first1} dM_{\mathrm{ther}}=T dS+\fft{(n-1)\omega}{256\pi (n+1)}\Big( (n-3)\psi d\mu_g+(n-1)\mu_g d\psi \Big) \,.\ee
The $\psi d\mu_g$ term on the r.h.s of this equation can be absorbed by defining a shifted mass
\be \widetilde{M}=M_{\mathrm{ther}}-\fft{(n-1)(n-3)\omega}{256\pi(n+1)}\psi\mu_g \,.\ee
The first law now looks more compact
\be\label{first2} d\widetilde{M}=T dS+\fft{(n-1)\omega}{128\pi(n+1)}\mu_g d\psi \,.\ee
Thus the solutions provide analytical examples how the first law of thermodynamics can be modified by a non-minimally coupled scalar in GB gravity. We have not found any analogous solutions in Einstein gravity non-minimally coupled to a scalar \cite{Fan:2015tua}.

The temperature and entropy are given by
\be T=\fft{(n-1)g^2r_0}{8\pi}\,,\qquad S=\fft 14 \omega r^{n-2}_0\Big(1-\fft{(n-1)}{8(n+1)}\phi^2_0  \Big) \,.\ee
where $\phi_0\equiv \phi(r_0)$. It is straightforward to verify that the thermodynamical first law (\ref{first1}) or (\ref{first2}) is indeed satisfied. The ``thermodynamic mass", temperature and entropy also satisfy a Smarr-like relation
\be M_{\mathrm{ther}}=\fft{n-2}{n-1}T S+\fft{(n-1)(n-2)\omega}{256\pi(n+1)}\psi\mu_g \,.\ee

\subsubsection{More solutions}
Inspired by above discussions, we may consider more general scalar potentials of the form
\be V=2\Lambda_0+\fft 12 m^2\phi^2+\gamma_4\phi^4 \,,\label{genev1}\ee
Indeed, we can construct more black hole solutions at the critical GB coupling $\alpha=\alpha_c$ provided that
\bea\label{genev2} &&\Lambda_0=-\fft 14(n-2)(n-1)g^2\,,\nn\\
&& m^2=\fft{\mu (n-2\mu-2)(n-2\mu-1)}{2(2\mu+1)}g^2\,,\nn\\
&& \gamma_4=-\fft{\mu^2(n-4\mu-2)(n-4\mu-1)}{32(2\mu+1)^2}g^2 \,.\eea
where $\mu$ is a reparametrization of $\eta$ in (\ref{phi}). The mass square of the scalar can be written as $m^2=m_{BF}^2+\ft 14(n-2\mu-1)^2g^2$, implying that the  BF bound is always satisfied. The black hole solutions read
\bea\label{ps3}
&&ds^2=-fdt^2+dr^2/f+r^2 dx^idx^i \,,\nn\\
&& \phi=\Big(\fft{q}{r}\Big)^{\mu}\,,\quad f=g^2r^2\Big(1-\fft{\mu q^{2\mu}}{4(2\mu+1)r^{2\mu}}\Big)\,.
\eea
The normalizability of the scalar at asymptotic infinity requires $\mu>0$ which also governs the existence of the event horizon.
 For $\mu=(n-1)/2$ and $\mu=(n-1)/4$, the solutions have been included as special cases in above two subsections. However, the global properties of the solutions are significantly different. We find that both the black hole mass and the entropy vanish for generic $\mu$, namely
\be \delta H_\infty=0 \,,\qquad S=0\,,\ee
although the temperature is non-vanishing $T=\fft{\mu r_0g^2}{2\pi}$. Thus the thermodynamical first law is trivially satisfied. Note that the black holes are
extremal at a finite temperature $T>0$, as well as the five dimensional spherical black hole (\ref{sol1}).

\subsubsection{Dynamical solutions}
For the potential (\ref{genev1}-\ref{genev2}), we can also obtain exact dynamical solutions when $\mu=(n-2)/2$. In Eddington-Finkelstein-like coordinates, the solutions read
\bea &&ds^2=-fdu^2-2dudr+r^2 dx^idx^i \,,\nn\\
     &&\phi=\Big(\fft{a}{r}\Big)^{n-2}\,,\qquad f=g^2r^2\Big(1-\fft{(n-2)a^{n-2}}{8(n-1)r^{n-2}}\Big)
\eea
where $a\equiv a(u)$ satisfies the second order non-linear differential equation (\ref{aeq}).
The new dynamical solutions are then simply given by $a=c_0/u$, where $c_0$ is a positive integration constant. The solutions describe a radiating white
hole at $u>0$ region and eventually become an AdS vacua at late retarded times.

\subsection{Charged hyperbolic black holes}

By introducing a Maxwell field $\mathcal{L}_A=-\ft 14 F^2$, we can obtain electrically charged black holes with hyperbolic topology for the potential given by
\bea V=&-&\fft 12(n-2)(n-1)g^2+\fft{(n-5)(n-4)(n-3)}{4(2n-5)}g^2\phi^2\nn\\
   &-&\fft{(3n-11)(3n-10)(n-3)^2}{32(2n-5)^2}g^2\phi^4 \,.\eea
The mass square of the scalar is given by $m^2=\fft{(n-5)(n-4)(n-3)}{2(2n-5)}g^2$, which is always above the BF bound for the non-minimal coupling $\eta=\fft{n-3}{2(2n-5)}$ because of $m^2=m_{BF}^2+\ft 14(n-5)^2g^2$. For this
non-minimal coupling and the critical GB coupling $\alpha=\alpha_c$, we obtain the black hole solutions
\bea ds^2&=&-fdt^2+dr^2/f+r^2 d\Omega_{n-2,-1}^2 \,,\quad \phi=\Big( \fft{q}{r}\Big)^{n-3} \,,\nn\\
      A&=&-\fft{q^{n-3}}{\sqrt{2n-5}r^{n-3}}dt\,,\qquad f=g^2 r^2-1-\fft{(n-3)g^2q^{2n-6}}{4(2n-5) r^{2n-8}} \,. \eea
By applying Wald formula, we find that the black hole mass vanishes $\delta H_\infty=0$. However, both the entropy and the temperature are non-vanishing
\be S=-\fft{\omega r^{n-4}_0}{2(n-4)g^2} \,,\qquad  T=\fft{g^2 r_0}{8\pi(2n-5)}\Big(4(2n-5)+(n-3)(n-4)\phi^2_0 \Big)\,.\ee
The conserved electric charge and the conjugate potential can be computed using standard
technique
\be Q\equiv\fft{1}{16\pi}\int {}^*F=\fft{(n-3)\omega q^{n-3}}{16\pi\sqrt{2n-5}} \,,\qquad \Phi_e\equiv A_t(\infty)-A_t(r_0)=\fft{q^{n-3}}{\sqrt{2n-5}r_0^{n-3}}\,.\ee
It follows that the thermodynamical first law reads
\be 0=T dS +\Phi_e dQ \,.\ee
The Smarr-like relation is given by
\be 0=T S+\Big( 1+\ft{4(2n-5)}{(n-3)(n-4)\phi_0^2} \Big)\Phi_e Q \,.\ee



\section{ Lifshitz black holes}
It was first shown in \cite{Dehghani:2010kd,Dehghani:2010gn} that for the critical GB coupling where there exists an unique AdS vacua, Lifshitz vacuum solutions
are allowed in GB gravity, namely
\be ds^2=\ell^2\Big(-r^{2z}dt^2+\fft{dr^2}{r^2}+r^2d\Omega_{n-2,k}\Big) \,,\qquad \alpha=\alpha_c=\fft{1}{2(n-3)(n-4)g^2}\,.\ee
 This motivates us to construct exact Lifshitz black hole solutions in our gravity model. We find that there are large classes of neutral and charged Lifshitz black holes with scalar hair for a either minimally or non-minimally coupled scalar with appropriate potentials.

\subsection{With a minimally coupled scalar}
Although we have not found any AdS black holes for a minimally coupled scalar, we obtain many new classes of Lifshitz black holes in this case. In fact, the potential and the solutions are much simpler than those constructed for a non-minimally coupled scalar.

\subsubsection{Neutral black holes}\label{mininbh}
Given the scalar potential
\be V=-\fft 12 (n-1)(n-2)g^2-\fft 12(n-1)^2g^2\phi^2-\fft{(n-1)^3}{8(n-2)}g^2\phi^4 \,,\ee
there exist exact black hole solutions for the Lifshitz exponent $z=n$
\bea &&ds^2=-r^{2n}\tilde{f}dt^2+\fft{dr^2}{r^2\tilde{f}}+r^2dx^idx^i \,,\nn\\
      && \phi=\Big(\fft{q}{r}\Big)^{n-1}\,,\quad \tilde{f}=1-\fft{(n-1)q^{2n-2}}{2(n-2)r^{2n-2}}\,,
\eea
where the effective AdS radius has been set to unity. The mass square of the scalar is given by $m^2=-(n-1)^2g^2$ which exactly equals to the BF bound\footnote{In asymptotically Lifshitz space-time, the Breitenlohner-Freedman (BF) bound for a non-minimally coupled scalar is given by $m^2_{BF}=-\fft{1}{4} (n+z-2)g^2+\eta \,g^2\Big(n^2+(2z-3)n+2(z-1)^2\Big)$.} at $z=n$.
The mass of the black holes can be read off from the Wald formula
\be M=\fft{(n-1)\omega q^{2n-2} }{32\pi}\,. \ee
The temperature and entropy are given by
\be T=\fft{(n-1)r_0^n}{2\pi}\,,\qquad S=\fft 14\omega r_0^{n-2}\,.\ee
It follows that the thermodynamical first law
\be dM=T dS \,,\ee
and the Smarr-like relation
\be M=\fft{n-2}{2(n-1)}T S \,,\label{nonsmarr}\ee
are satisfied. It is worth pointing out that for Lifshitz planar black holes, one can derive the Smarr-like relation by studying the global scaling symmetry of the space-time \cite{Hyun:2015tia}.

\subsubsection{Charged black holes}\label{minicbh}
Introducing an additional Maxwell field, we obtain two different classes of charged Lifshitz black hole solutions. The first class solution has the potential of
\be V=-\fft 12(n-1)(n-2)g^2-\fft 12(n-1)(n-2)g^2\phi^2-\fft 18(n-2)(n-3)g^2\phi^4 \,.\ee
The mass square of the scalar is given by $m^2=-(n-1)(n-2)g^2$ which does not satisfy the BF bound for generic $z$. Fortunately, for the Lifshitz exponent $z=n-1$ where we find the black hole solutions, the BF bound is always satisfied because of $m^2=m_{BF}^2+\ft 14 g^2$. The black hole solutions read
\bea &&ds^2=-r^{2n-2}\tilde{f}dt^2+\fft{dr^2}{r^2\tilde{f}}+r^2dx^idx^i\,,\quad \phi=\Big(\fft{q}{r}\Big)^{n-2} \,,\nn\\
      &&A=\sqrt{n-2}\,q^{n-2}r dt\,,\qquad \tilde{f}=1-\fft{q^{2n-4}}{2r^{2n-4}}\,,
\eea
The Wald formula implies that the black hole mass vanishes $\delta H_\infty=0$. The electric charge can be calculated using standard technique
\be Q=\fft{\sqrt{n-2}\,\omega q^{n-2}}{16\pi}\,, \ee
while the common definition of the conjugate potential breaks down since the gauge potential diverges at asymptotic infinity. This should be treated carefully. A refined definition for the electric potential has been provided in \cite{Fan:2014ala,Liu:2014dva,Fan:2015yza,Fan:2015aia} using Wald formalism
\be \Phi_e\equiv A^{\mathrm{reg}}_t(\infty)-A_t(r_0)\,,\qquad A^{\mathrm{reg}}_t(\infty)= A_t(\infty)-A^{\mathrm{div}}_t(\infty) \,.\ee
For our solutions, we find
\be \Phi_e=-\sqrt{n-2}\,q^{n-2}r_0 \,.\ee
The temperature and entropy are given by
\be
T=\fft{(n-2)r_0^{n-1}}{2\pi}\,,\qquad S=\fft 14\omega r_0^{n-2}\,.
\ee
It follows that the thermodynamical first law
\be 0=T dS+\Phi_e dQ \,,\label{lifc1}\ee
and the Smarr-like relation
\be 0=T S+\Phi_e Q \,,\label{lifsmarr1}\ee
straightforwardly hold. The first law of this type for charged Lifshitz black holes was first studied in \cite{Liu:2014dva} using Wald formalism.

The second class solution has the potential of
\be V=-\fft 12(n-1)(n-2)g^2-\fft 32 (n-2)^2g^2\phi^2+\fft{(n-2)^3}{8(3n-7)}g^2\phi^4 \,.\ee
The scalar mass square is given by $m^2=-3(n-2)^2g^2$ and the BF bound is always satisfied for the exponent $z=3(n-2)$ where we find the black hole solutions owing to the relation $m^2=m_{BF}^2+(n-2)^2g^2$. The solutions read
\bea &&ds^2=-r^{6n-12}\tilde{f}dt^2+\fft{dr^2}{r^2\tilde{f}}+r^2dx^idx^i\,,\quad \phi=\Big(\fft{q}{r}\Big)^{n-2} \,,\nn\\
      &&A=\fft{1}{\sqrt{2}}\,q^{n-2}r^{2n-4} dt\,,\qquad \tilde{f}=1-\fft{(n-2)q^{2n-4}}{2(3n-7)r^{2n-4}}\,,
\eea
We find that the black hole mass is proportional to the coefficient square of the fall-off mode $1/r^{2n-4}$, given by
\be M=-\fft{(n-2)^2(2n-5)\omega q^{4n-8}}{128\pi(3n-7)^2} \,.\ee
Other thermodynamic quantities are given by
\bea
&&T=\fft{(n-2)r_0^{3n-6}}{2\pi}\,,\qquad S=\fft 14 \omega r_0^{n-2}\,,\nn\\
&&\Phi_e=-\fft{q^{n-2}r_0^{2n-4}}{\sqrt{2}}\,,\qquad Q=\fft{(n-2)\omega q^{n-2}}{8\sqrt{2}\pi}\,.
\eea
It follows that the thermodynamical first law reads
\be dM=T dS+\Phi_e dQ \,.\label{lifc2}\ee
The Smarr-like relation turns out to be underdetermined, given by
\be M=\beta_1\, T S+\beta_2\, \Phi_e Q\,,\qquad (n-2)\beta_1-(3n-7)\beta_2+\ft 14(2n-5)=0 \,.\label{lifsmarr2}\ee
This is not a surprise since the solutions are highly degenerated.

\subsection{With a non-minimally coupled scalar}

\subsubsection{Neutral black holes}
For a non-minimally coupled scalar field, we obtain two different classes of solutions. The first class solution generalizes the minimal solutions in sec \ref{mininbh}. The potential is given by
\bea V&=&-\fft 12(n-1)(n-2)g^2-\fft 12(n-1)\Big(n-1-(5n-2)\eta \Big)g^2\phi^2 \nn\\
 &&-\fft{n-1}{8(n-2)}\Big((\eta-1)(5\eta-1)n^2-2(2\eta-1)(3\eta-1)n+(2\eta-1)^2  \Big)  g^2\phi^4 \,.\eea
The black hole solutions exist for $z=n$
\bea &&ds^2=-r^{2n}\tilde{f}dt^2+\fft{dr^2}{r^2\tilde{f}}+r^2dx^idx^i \,,\nn\\
      && \phi=\Big(\fft{q}{r}\Big)^{n-1}\,,\quad \tilde{f}=1-\fft{\big(n-1-(5n-2)\eta \big)q^{2n-2}}{2(n-2)r^{2n-2}}\,,
\eea
The existence of the event horizon requires $\eta<(n-1)/(5n-2)$. Hence, the scalar has a negative mass square $m^2=-(n-1)\Big(n-1-(5n-2)\eta \Big)g^2$, which exactly equals to the BF bound (for $z=n$). The black hole mass is given by
\be  M=\fft{\big(n-1-2(3n-2)\eta \big)\omega q^{2n-2} }{32\pi}\,. \ee
The temperature and entropy are given by
\be T=\fft{(n-1)r_0^n}{2\pi}\,,\qquad S=\fft 14\omega r_0^{n-2}\Big(1-\ft 12\eta \phi_0^2\Big)\,.\ee
It is straightforward to verify that the thermodynamical first law $dM=T dS$ and the Smarr-like relation (\ref{nonsmarr}) are satisfied.

The second class solution we present has a new potential
\bea\label{v3} V&=&-\fft 12(n-1)(n-2)g^2+\fft{\mu}{4(2\mu+z)}\Big(n^2-(4\mu+3)n+2\big(2\mu^2-(z-4)\mu+1\big) \Big) g^2\phi^2 \nn\\
       &&-\fft{\mu^2}{32(2\mu+z)^2}\Big(n^2-(8\mu+3)n+2\big( 8\mu^2-2(z-4)\mu+1 \big) \Big) g^2\phi^4 \,.
\eea
The scalar always satisfies the BF bound because of $m^2=m_{BF}^2+\ft 14(n+z-2\mu-2)^2g^2$.
The black hole solutions exist for generic exponent $z$, given by
\bea &&ds^2=-r^{2z}\tilde{f}dt^2+\fft{dr^2}{r^2\tilde{f}}+r^2dx^idx^i \,,\nn\\
      && \phi=\Big(\fft{q}{r}\Big)^{\mu}\,,\qquad \tilde{f}=1-\fft{\eta\, q^{2\mu}}{2r^{2\mu}}\,,
\eea
where the non-minimal coupling constant is determined by
\be \eta=\fft{\mu}{2(2\mu+z)} \,.\ee
The existence of the event horizon requires $\eta>0$ and hence $z>-2\mu$. It turns out that the black holes have both vanishing mass and entropy. Thus the thermodynamical first law is trivially satisfied.

\subsubsection{Dynamical solutions}
When $\mu=(n-2)/2$, the potential (\ref{v3}) becomes simplified
\be V=-\fft 12 (n-1)(n-2)g^2-\fft{(n-2)^2(z-1)}{8(n+z-2)}g^2\phi^2-\fft{(n-2)^3(n-2z-1)}{128(n+z-2)^2}g^2\phi^4 \,.\ee
We obtain exact dynamical Lifshitz solutions in Eddington-Finkelstein-like coordinates. The solutions read
\bea
&&ds^2=-r^{2z}\tilde{f}du^2-2r^{z-1}dr du+r^2 dx^idx^i\,,\nn\\
&&\phi=\Big(\fft{a}{r} \Big)^{\fft{n-2}{2}}\,,\qquad \tilde{f}=1-\fft{(n-2)a^{n-2}}{8(n+z-2)r^{n-2}}\,.
\eea
The time dependent ``scalar charge" $a$ satisfies a second order non-linear differential equation
\be a\ddot{a}-(z+1)\dot{a^2}=0 \,,\ee
which can be immediately solved by
\be a=q\,,\qquad \mathrm{or}\qquad a=\fft{c_0}{u^{1/z}}\,, \ee
where $c_0$ is a positive integration constant. The new dynamical solutions describe radiating white holes which approach the Lifshitz vacuum at late retarded times.

\subsubsection{Charged black holes}

Introducing a Maxwell field, we find that the charged Lifshitz black holes in sec \ref{minicbh} for a minimally coupled scalar can be properly generalized for a non-minimal coupled scalar. The first class solution
we present has the potential of
\bea V&=&-\fft 12 (n-1)(n-2)g^2-\fft 12(n-1)\big(n-1-(5n-8)\eta \big) g^2\phi^2 \nn\\
      &&-\fft{1}{8(n-2)}\Big((\eta-1)(5\eta-1)n^3-(\eta-1)(33\eta-7)n^2\nn\\
      &&+4(15\eta^2-21\eta+4)n-4(2\eta-3)(4\eta-1) \Big)g^2\phi^4\,.
\eea
The black hole solutions read
\bea &&ds^2=-r^{2n-2}\tilde{f}dt^2+\fft{dr^2}{r^2\tilde{f}}+r^2dx^idx^i\,,\quad \phi=\Big(\fft{q}{r}\Big)^{n-2} \,,\nn\\
      &&A=\sqrt{n-2-2(3n-5)\eta}\,q^{n-2}r dt\,,\qquad \tilde{f}=1-\fft{\big(n-2-(5n-8)\eta\big)q^{2n-4}}{2(n-2)r^{2n-4}}\,,
\eea
Note that the BF bound for $z=n-1$ is always satisfied because of $m^2=m_{BF}^2+\ft 14 g^2$. Stability of the solutions requires $\eta<(n-2)/\big(2(3n-5)\big)$, which also governs the existence of the event horizon. We find that the black hole mass vanishes $\delta H_\infty=0$. Other thermodynamical quantities are given by
\bea
&&T=\fft{(n-2)r_0^{n-1}}{2\pi}\,,\qquad S=\fft 14\omega r_0^{n-2}\Big(1-\fft{(n-2)\eta}{n-2-(5n-8)\eta}\Big)\,,\nn\\
&&\Phi_e=-\sqrt{n-2-2(3n-5)\eta}\,q^{n-2}r_0\,,\qquad Q=\fft{\sqrt{n-2-2(3n-5)\eta}\,\omega q^{n-2}}{16\pi}\,.
\eea
It is straightforward to verify that the thermodynamical first law (\ref{lifc1}) and the Smarr-like relation (\ref{lifsmarr1}) are satisfied.

The second class solution we present has the potential of
\bea V&=&-\fft 12(n-1)(n-2)g^2-\fft 12 (n-2)\Big(3(n-2)-(25n-49)\eta \Big)g^2\phi^2\nn\\
&&+\fft{(n-2)}{8(3n-7)}\Big(n-2-(7n-13)\eta\Big)^2g^2\phi^4 \,.\eea
The black hole solutions read
\bea &&ds^2=-r^{6n-12}\tilde{f}dt^2+\fft{dr^2}{r^2\tilde{f}}+r^2dx^idx^i\,,\quad \phi=\Big(\fft{q}{r}\Big)^{n-2} \,,\nn\\
      &&A=\sqrt{\ft 12-5\eta}\,q^{n-2}r^{2n-4} dt\,,\qquad \tilde{f}=1-\fft{\big(n-2-(7n-13)\eta \big)q^{2n-4}}{2(3n-7)r^{2n-4}}\,,
\eea
The BF bound for the exponent $z=3(n-2)$ is always satisfied because of $m^2=m_{BF}^2+(n-2)^2g^2$. Stability of the solutions requires $\eta<1/10$, which also governs the existence of the event horizon. The black hole mass turns out to be non-vanishing, given by
\be M=\fft{(n-2)(2n-5)(10\eta-1)\Big(n-2-(7n-13)\eta\Big)\omega q^{4n-8}}{128\pi(3n-7)^2} \,.\ee
Some other quantities are given by
\bea
&&T=\fft{(n-2)r_0^{3n-6}}{2\pi}\,,\qquad S=\fft 14 \omega r_0^{n-2}\Big(1-\fft{(3n-7)\eta}{n-2-(7n-13)\eta} \Big)\,,\nn\\
&&\Phi_e=-\sqrt{\ft 12-5\eta}\,q^{n-2}r_0^{2n-4}\,,\qquad Q=\fft{(n-2)\sqrt{1-10\eta}\, \omega q^{n-2}}{8\sqrt{2}\pi}\,.
\eea
It follows that the thermodynamical first law (\ref{lifc2}) and the Smarr-like relation (\ref{lifsmarr2}) are satisfied.

\section{Conclusion}
In this paper, we study Gauss-Bonnet gravity in general dimensions non-minimally coupled to a scalar field. In order to construct exact black
hole solutions as many as possible, we also introduce a minimally coupled Maxwell field. With an appropriate scalar potential of the type
$V=2\Lambda_0+\fft 12 m^2\phi^2+\gamma_4 \phi^4$, we construct large classes of AdS and Lifshitz black hole solutions with scalar hair.

First, we obtain a spherical black hole in five dimension, which is asymptotic to locally AdS space-times. We then obtain three classes of AdS planar black holes
and one class of AdS hyperbolic black holes with electric charges in general dimensions. We study the global properties of the solutions and derive the
thermodynamical first laws using Wald formalism. For most of our solutions except for one class of the planar black holes, there is only one independent integration constant $q$ associated
with the scalar and all the thermodynamical quantities are determined by functions of $q$. It turns out that the spherical black hole as well as one class of planar black holes
has both vanishing mass and entropy and the first law is trivially satisfied. The charged hyperbolic black holes have vanishing mass and but non-vanishing
entropy and the first law $0=TdS+\Phi_e dQ$ is satisfied. In particular, one class of the planar black hole solutions contains two independent integration constants,
which are associated with the scalar hair and the graviton mode respectively. Although the scalar has only one normalizable mode at asymptotic infinity,
it forms a thermodynamical conjugate with the graviton mode and non-trivially contributes to the first law of thermodynamics.
The solutions provide analytical examples how the thermodynamical first laws can be modified by a non-minimally coupled scalar in GB gravity.

We observe that except for one class of the solutions, all the AdS solutions constructed above are valid at the critical point of GB gravity where there
exists an unique AdS vacua. In fact, more general Lifshitz vacua is also allowed at this point. We then try to construct exact Lifshitz black hole solutions with
scalar hair in our gravity model. We obtain many new classes of neutral and electrically charged Lifshitz black holes for a either minimally or
non-minimally coupled scalar and derive the thermodynamical first laws.

 Finally, for some classes of the solutions such as the spherical black hole, one class of AdS planar black holes and one class of Lifshitz black holes, we
 promote the ``scalar charge" to be dependent on the retarded time and obtain exact dynamical solutions. The solutions describe radiating white holes and
 eventually become the AdS or Lifshitz vacua at late retarded times. However, the spherical black hole approaches a static AdS space-time with a globally naked singularity.

\section*{Acknowledgments}
The authors thank Prof. Hong Lu for valuable discussions. This work was in part supported by NSFC Grants No.~11275010, No.~11335012 and No.~11325522.

\section{Appendix: An Explicit Example for Large-r Expansion}
We assume the scalar potential has
a stationary point at $\phi=0$ and can be expanded around it as
\be V=2\Lambda_0+\fft 12 m^2\phi^2+\gamma_3\phi^3+\gamma_4\phi^4+\cdots \,,\ee
where $\Lambda_0$ is the bare cosmological constant and $m$ is the mass of the scalar. At asymptotically AdS space-time, the scalar field has two independent
fall-offs
\be \phi=\fft{\psi_1}{r^{\fft{n-1-\sigma}{2}}}+\fft{\psi_2}{r^{\fft{n-1+\sigma}{2}}}+\cdots \,,\label{fallscalar}\ee
where $\sigma$ is defined by
\be \sigma=\sqrt{(n-1)\Big((1-4\eta)n-1\Big)+4m^2\ell^2} \,.\ee
The full large-r expansion of the general
static solutions strongly depend on the value of the scalar mass and the coupling constants for our gravity model. For simplicity, we focus on $0<\sigma<1$ case, corresponding to the scalar mass
\be m^2_{BF}< m^2<m^2_{BF}+1 \,.\ee
It is clear that the leading fall-off of the scalar is given by (\ref{fallscalar}) but there could be slower fall-offs than those displayed for general scalar mass parameter. For generic coupling constants, the asymptotical behavior of the metric functions are of the forms
\be h=g^2 r^2+k-\fft{\mu_g}{r^{n-3}}+\cdots\,,\qquad f=g^2 r^2+k+\fft{a}{r^{n-3-\sigma}}-\fft{\tilde{\mu}_g}{r^{n-3}}+\cdots \,,\ee
where $\mu_g$ is associated with the graviton mode and
\bea &&a=\fft{(n-1)\Big((1-4\eta)(n-\sigma)-1\Big)g^2\psi_1^2}{4(n-1)(n-2)\Big(1-2\alpha g^2(n-3)(n-4)\Big)}\,,\nn\\
&&\tilde{\mu}_g=\mu_g-\fft{\Big((n-1)^2-4\eta n(n-1)-\sigma^2\Big)g^2\psi_1\psi_2}{2(n-1)(n-2)\Big(1-2\alpha g^2(n-3)(n-4)\Big)} \,.
\eea
Substitute the large-r expansion into the Wald formula, we obtain
\bea\label{examplek} \delta H_{\infty}&=&\fft{(n-2)\omega}{16\pi}\Big(1-2\alpha g^2(n-3)(n-4)\Big)\delta \mu_g+\nn\\
&&\fft{\sigma\omega g^2}{32\pi(n-1)}\Big((n+\sigma-1)\psi_2\delta \psi_1
-(n-\sigma-1)\psi_1\delta \psi_2  \Big) \,.\eea
Thus the thermodynamical first law can be expressed as
\be dM_{\mathrm{ther}}=T dS-\fft{\sigma\omega g^2}{32\pi(n-1)}\Big((n+\sigma-1)\psi_2\delta \psi_1
-(n-\sigma-1)\psi_1\delta \psi_2  \Big) \,,\ee
where the ``thermodynamic mass" is defined by
\be M_{\mathrm{ther}}=\fft{(n-2)\omega}{16\pi}\Big(1-2\alpha g^2(n-3)(n-4)\Big) \mu_g \,,\ee
which is consistent with the generalized  AMD mass in higher curvature gravity \cite{Okuyama:2005fg,Pang:2011cs}. Note that the one form associated with the scalar vanishes in (\ref{examplek}) if the scalar has only one normalizable mode at asymptotic infinity or the two independent fall-offs of the scalar satisfy an identity $\psi_2^{\Delta_1}\propto \psi_1^{\Delta_2}$, where $\Delta_1=(n-1-\sigma)/2\,,\Delta_2=(n-1+\sigma)/2$ are the scaling dimensions associated with $\psi_1$ and $\psi_2$ respectively. This is also true for generic parameters.

\end{document}